\documentclass[journal]{IEEEtran}

\IEEEoverridecommandlockouts   
\usepackage[T1]{fontenc}
\usepackage{url}
\usepackage{doi}

\usepackage{xcolor}
\usepackage{hyperref}
\usepackage{graphicx}
\usepackage{float}
\usepackage{adjustbox}
\usepackage{rotating}
\usepackage{listings}
\usepackage{booktabs}
\usepackage{multirow}
\usepackage{makecell}
\usepackage{threeparttable}   
\usepackage{array}
\usepackage{diagbox}
\usepackage{caption}
\graphicspath{{./images/} {../pdf/}{../jpeg/}}
\DeclareGraphicsExtensions{.pdf,.jpeg,.png}

\usepackage{amsmath}
\usepackage{mathtools}
\usepackage{amssymb}
\usepackage{mdwmath}
\usepackage{mdwtab}
\usepackage{eqparbox}
\usepackage{breqn}
\usepackage{physics}

\usepackage[nocompress,space]{cite}

\usepackage[ruled,vlined]{algorithm2e}

\usepackage{pdfrender}
\usepackage{soul,color}
\usepackage{orcidlink}
\usepackage{tikz} 
 
\usepackage{pifont}  
\usepackage{bbding}
\usepackage{wasysym}

\usepackage[
  separate-uncertainty = true,
  multi-part-units = repeat
]{siunitx}
\usepackage{textcomp}

\usepackage{cleveref}

\usepackage{blindtext}
\usepackage{lipsum}
\usepackage{comment}

\newif\ifshowmods
\showmodstrue     

\ifshowmods

  \newcommand{\Ablack}[1]{\textcolor{black}{#1}} 
\else

  \newcommand{\Ablack}[1]{#1}
\fi

\begin{document}

\title{

Large Language Models for Detecting Cyberattacks on Smart Grid Protective Relays

}

\author{
Ahmad~Mohammad~Saber\orcidlink{0000-0003-3115-2384},~\IEEEmembership{Member,~IEEE,} 
Saeed Jafari\orcidlink{0009-0006-0516-8605},
Zhengmao Ouyang\orcidlink{0009-0008-7875-0465},
Paul Budnarain\orcidlink{0009-0006-5638-2310},
Amr~Youssef\orcidlink{0000-0002-4284-8646},~\IEEEmembership{Senior Member,~IEEE,} 
and~Deepa~Kundur\orcidlink{0000-0001-5999-1847},~\IEEEmembership{Fellow,~IEEE}

\thanks{
Ahmad Mohammad Saber, 
Saeed Jafari, 
Zhengmao Ouyang, Paul Budnarain, and Deepa Kundur are with 
the Department of Electrical and Computer Engineering, University of Toronto, Toronto, ON M5S 1A1, Canada
(e-mails: \href{mailto:ahmad.m.saber@ieee.org}{ahmad.m.saber@ieee.org}; 
\href{mailto:saeed.jafari@mail.utoronto.ca}{saeed.jafari@mail.utoronto.ca};
\href{mailto:zhengmao.ouyang@mail.utoronto.ca}{zhengmao.ouyang@mail.utoronto.ca};
\href{mailto:paul.budnarain@mail.utoronto.ca}{paul.budnarain@mail.utoronto.ca};
\href{mailto:dkundur@ece.utoronto.ca}{dkundur@ece.utoronto.ca})}
\thanks{Amr Youssef is with the Concordia Institute for Information Systems Engineering (CIISE), Concordia University, Montreal, QC, Canada
(e-mail: 
\href{mailto:youssef@ciise.concordia.ca}{youssef@ciise.concordia.ca}).}
\thanks{The first four authors contributed equally to the conceptualization and overall preparation of this manuscript.}

}

\markboth{
}%
{Shell \MakeLowercase{\textit{et al.}}: Bare Demo of IEEEtran.cls for IEEE Journals}

\maketitle

\begin{abstract}

This paper presents a large language model (LLM)–based framework that adapts and fine-tunes compact LLMs for detecting cyberattacks on transformer current differential relays (TCDRs), which can otherwise cause false tripping of critical power transformers. The core idea is to textualize multivariate time-series current measurements from TCDRs, across phases and input/output sides, into structured natural-language prompts that are then processed by compact, locally deployable LLMs. Using this representation, we fine-tune DistilBERT, GPT-2, and DistilBERT+LoRA to distinguish cyberattacks from genuine fault-induced disturbances while preserving relay dependability. The proposed framework is evaluated against a broad set of state-of-the-art machine learning and deep learning baselines under nominal conditions, complex cyberattack scenarios, and measurement noise. Our results show that LLM-based detectors achieve competitive or superior cyberattack detection performance, with DistilBERT detecting up to 97.62\% of attacks while maintaining perfect fault detection accuracy. Additional evaluations demonstrate robustness to prompt formulation variations, resilience under combined time-synchronization and false-data injection attacks, and stable performance under realistic measurement noise levels. The attention mechanisms of LLMs further enable intrinsic interpretability by highlighting the most influential time–phase regions of relay measurements. These results demonstrate that compact LLMs provide a practical, interpretable, and robust solution for enhancing cyberattack detection in modern digital substations. We provide the full dataset used in this study for reproducibility. 
 
\end{abstract}


\IEEEpeerreviewmaketitle

\section{Introduction}\label{section:Introduction}

\IEEEPARstart{M}{odern} substations leverage networked devices and real-time data exchange to enhance protection and control; however, their increasing dependence on advanced information and communication technologies also introduces significant cybersecurity vulnerabilities across critical substation components \cite{10065529}.
Power transformer current differential relays (TCDRs), common to these substations, rely on communicated measurements, making them vulnerable to cyberattacks aiming to falsely trip  power transformers.
Detecting such cyberattacks requires distinguishing manipulated TCDR measurements from those under faults while preserving the TCDR's fault detection capability. 

Large language models (LLMs) represent a breakthrough in AI offering high accuracy and interpretability in processing structured textual data \cite{sanh2019distilbert}. 
\Ablack{Recent studies have introduced LLMs to some smart grid applications. Previous efforts have explored LLMs for tasks such as  partial \Ablack{distributed energy resource} tripping identification \cite{zhao2024large}, anomaly detection in communication protocols \cite{zaboli2024chatgpt}, and non-intrusive load monitoring \cite{11245588}. 
Work \cite{zaboli2024chatgpt} leveraged cloud-based LLMs and the data was primarily textual. In \cite{zhao2024large}, data was numerical but with very low-dimensional feature sets. 
The scheme in \cite{11245588} relies on cloud-based LLMs. 
The scope of these works, data characteristics, and deployment models differ substantially from the approach presented in this work.
This work investigates the use of LLMs for detecting false data injection attacks (FDIAs) against transformer current differential relays (TCDRs), a cyber-physical protection challenge that has not been addressed in prior literature. By focusing on high-resolution relay measurements and their contextual interpretation, our framework introduces a new use case for LLMs in the domain of power system protection and resilience.}

This paper demonstrates the effectiveness of LLMs in detecting cyberattacks on TCDRs using only current measurements, adding a strong layer of security to smart grids. This work contributes to power system protection in three key ways:
\begin{itemize}
    \item We develop a method to convert current measurements (from all six input and output phases) into structured text prompts. This approach captures the physical relationships needed for fault analysis while fitting within the LLM's input size limits. 
    \item We prove that lightweight LLMs (DistilBERT, GPT-2) can be deployed locally within digital substations. We show that these models operate in less than 6 ms on standard commercial hardware, meeting the strict timing requirements of protective relays without needing external cloud infrastructure. 
    \item We introduce a transparent detection method for False Data Injection Attacks (FDIAs) using the model's self-attention mechanism. Unlike `black-box' models, this approach allows protection engineers to clearly see which time steps and phases show signs of an attack, thereby improving trust in the automated decision.
\end{itemize}
First, we adapt DistilBERT, a lightweight LLM that can be deployed locally, to distinguish between manipulated TCDR measurements and those caused by actual faults. This is achieved by fine-tuning the model on various cyberattack and of fault cases, where TCDR measurements of each case are textualized  using a standard prompt.
Additionally, we implement LLMs GPT-2 and DistilBERT+LoRA, as well as several deep learning (DL) and machine learning (ML) models, for performance comparison. 
Further, we  demonstrate  that the proposed LLM assigns high attention score to relay measurements affected by the cyberattack or fault immediately after the event's unfolding. 
\Ablack{Our results confirm that the proposed framework detects 97.6\% of cyberattacks while ensuring the relay remains reliable during actual faults, outperforming state of the art models. Furthermore,  our results also demonstrate the robustness of the proposed LLM-based framework under complex attacks and measurement noise, its stability across variations in the prompt template, and real-time deployability with an inference time of less than 6 ms on a commercial PC.}


\section{Cyberattacks Against TCDRs }\label{section:problem}

\begin{figure}[t!]
\centering
\includegraphics[width=1\columnwidth]{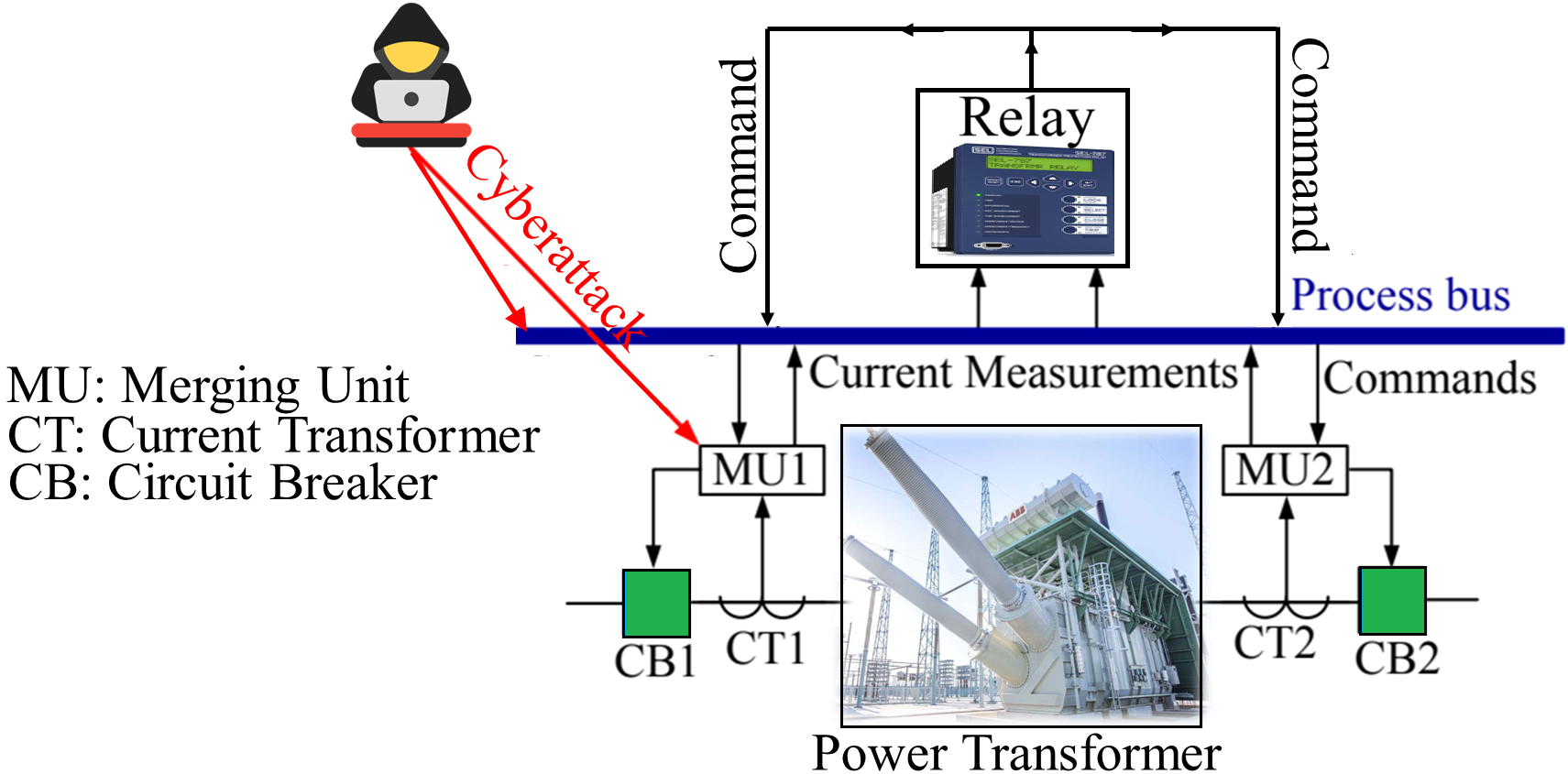}
\caption{Possible intrusion points for cyberattacks on a TCDR.}
\label{fig:Vulnerability}
\end{figure}

Existing TCDRs' logic cannot distinguish between fault-induced measurements and those manipulated during a false-tripping cyberattack. TCDRs are designed to detect internal transformer faults by comparing the phasors of the transformer's 3-phase phase current measurements from both its input and output sides  \cite{SEL_inc}.  Under normal conditions, these measurements are nearly identical, but faults cause significant differences in magnitude or phase angle. However, as shown in Fig. \ref{fig:Vulnerability}, malicious actors can exploit vulnerabilities in the communication protocols on which modern IEC-61850-based substations hosting TCDRs are designed \cite{10065529}. 
This could enable the malicious entities to manipulate transmitted TCDR measurements, fooling the TCDR into unnecessary tripping of its power transformer under normal system operation, leading to prolonged outages and potential grid instability in case of coordinated attacks \cite{10065529,khaw2020deep}. 

Detecting such cyberattacks requires a solution capable of analyzing the TCDR's six measurements before and after the TCDR is triggered to trip, to distinguish between malicious and legitimate fault measurements. 
The challenge lies in handling high-dimensional, complex data while ensuring accurate cyberattack detection without compromising TCDR fault detection accuracy.

\section{ LLMs for Detecting Cyberattacks on TCDRs}

The proposed LLM-based detection framework contextualizes TCDR measurements using a standard template prompt, which is then processed by a fine-tuned LLM to validate their authenticity, acting as a cyberattack detector for each critical TCDR.
Once deployed within the substation, the framework continuously monitors incoming TCDR measurements. When the TCDR is triggered, whether by a fault or a cyberattack, the LLM-based framework assesses the situation. If a potential attack is detected, the TCDR’s tripping command to the circuit breakers can be blocked and the operator is alerted.
The first step in developing this framework involves designing a standard template to convert TCDR measurements into a structured textual format. Next, the LLM is fine-tuned on various attack and fault scenarios, learning to analyze textualized TCDR measurements, as explained below. 

\Ablack{It is important to emphasize that in this paper, the LLMs, e.g., DistilBERT, GPT-2, are not trained from scratch, which would require prohibitive computational resources. Instead, we fine-tune compact pretrained models using our domain-specific dataset. As will be explained in the upcoming sections, fine-tuning can be performed on commercial PCs, without the need for cloud-scale infrastructure. This approach ensures that the method remains both resource-efficient and practical for deployment in substation environments.}

\subsection{Contextualization of TCDR Measurements for LLMs}
TCDR measurements before and after triggering can be recorded, forming a six-dimensional sequence of numerical values. This measurement sequence is then contextualized using the following prompt:  

\begin{lstlisting}[basicstyle=\ttfamily\footnotesize, breaklines=true]
"Transformer Differential Relay's current measurement vector of phase A on transformer input side: [numerical values]
. . .
Transformer Differential Relay's current measurement vector of phase C on transformer output side: [numerical values]"
\end{lstlisting}

\noindent where \texttt{[numerical values]} represents the one-dimensional vector of sampled current measurements for the respective phase.  
In this prompt, structured text is incorporated before each measurement vector to  help the LLM interpret phase relationships between TCDR measurements as well as their locations on the transformer's input and output sides. 
This guides the LLM, with its prior knowledge, to learn the relationships in TCDR measurements, e.g., inter-phase dependencies and temporal patterns, under cyberattacks vs true faults.
The LLM process inputs as sequences of tokens (e.g., word, subword, decimal point, punctuation, or number).
To ensure consistent tokenization, all numerical inputs are preformated into uniformly fixed-length strings using a fixed number of decimal places in addition to normalization of the six-dimensional TCDR measurement sequence. In our approach, we use 512 tokens per sample, which is a number of tokens that can be supported by DistilBERT \cite{sanh2019distilbert}. \Ablack{The above representation procedure represents a key methodological contribution of this work, which enables sampled values to be encoded into structured prompts that preserve temporal physical dependencies while remaining within the token constraints of compact LLMs.}

\subsection{DistilBERT for Cyberattack Detection}
\subsubsection{DistilBERT as a sequence classifier}
DistilBERT is a lightweight variant of the BERT model, designed using knowledge distillation to balance sequence processing accuracy and computational efficiency \cite{sanh2019distilbert}. It is based on the Transformer architecture, which employs self-attention mechanisms to capture dependencies within sequential data.
Given an input sequence of $n$ tokens, \( X = (x_1, x_2, \dots, x_n) \), representing the tokenized textual form of TCDR measurements under a fault or cyberattack, the self-attention mechanism computes a weighted sum of token embeddings, where the weights reflect the relevance of each token to the others.
\Ablack{It is important to note that weights are not manually assigned to detect attacks. The detection task is performed by the fine-tuned LLM, which learns directly from the training data and outputs an attack/fault classification through its dense output layer. The ‘weights’ refer to the internal self-attention scores produced by the Transformer, where for each token pair \(i,j\), the attention score \( A_{ij} \) between tokens \( x_i \) and \( x_j \) is computed as follows \cite{sanh2019distilbert}: 
\begin{equation}
A_{ij}=\frac{\exp(Q_i \cdot K_j)}{\sum_{k}\exp(Q_i \cdot K_k)}\label{eq:1}
\end{equation}
\noindent where \( Q_i \) and \( K_j \) are the query and key vectors associated with tokens \( x_i \) and \( x_j \), respectively. These vectors are learned during training, enabling the model to identify patterns in TCDR measurements related to faults and cyberattacks.
These attention scores are generated automatically as part of the model’s inference process and can be averaged across heads and layers to obtain token-level importance scores. By projecting these scores back to the measurement domain, we highlight influential time–phase regions of the current waveforms. This provides interpretability for power system operators, while the ultimate decision (attack or fault) remains the direct output of the fine-tuned LLM's classification head.}

The proposed model employs multiple layers, each with several attention heads, allowing it to detect distinguishing features in TCDR measurements, such as temporal variations in current magnitudes and phase angles. Each attention head applies the scaled dot-product attention from Equation \ref{eq:1} using learned weight matrices for \( Q \), \( K \), and \( V \). The outputs from all heads are then concatenated and passed through a linear transformation. 
\Ablack{Finally, the leveraged LLM encompasses a fully connected layer for binary classification, 
enabling binary prediction of the input event as either a fault or an FDIA.}

\subsubsection{Fine-Tuning DistilBERT}

Fine-tuning a pre-trained LLM involves updating a subset of its parameters, including the weights of classification layer, to adapt the LLM to cyberattack detection. This approach reduces computational overhead while leveraging the model’s pre-trained knowledge.
The LLM is fine-tuned on tokenized textual representations of TCDR measurements, learning to classify each sequence as either a fault or a cyberattack \cite{zhao2024large}.
Moreover, attention mechanisms provide a kind of interpretability by identifying the most influential tokens on the LLM's decision. Token-level attention scores, derived from averaged attention weights across multiple layers, highlight the relative importance of each token to the model’s decision. 
This enables the identification of the most influential input tokens and their corresponding time steps, offering insights into the model's decisions \cite{sanh2019distilbert}.


\begin{table*}[t!]
\centering
\caption{Performance Evaluation Results (Main Case Study)}
\label{tab:weighted_metrics}
\renewcommand{\arraystretch}{1.25}
\begin{tabular}{>{\centering\arraybackslash}m{2em} l c c c c c c}
\Xhline{3\arrayrulewidth}
\rotatebox{90}{\textbf{ }} & \textbf{Model} & \textbf{Cyberattack Detection Rate (\%)} & \textbf{Accuracy (\%)} & \textbf{Precision (\%)} & \textbf{Recall (\%)} & \textbf{Specificity (\%)} & \textbf{F1-Score (\%)} \\
\Xhline{3\arrayrulewidth}
\rule{0pt}{2ex} \multirow{3}{*}{\rotatebox{90}{LLMs}}
& \textbf{DistilBERT} & \textbf{97.62} & \textbf{99.84} & \textbf{100.00} & \textbf{98.81} & \textbf{100.00} & \textbf{99.36} \\
& GPT-2 & 97.06 & 99.80 & 99.90 & 98.53 & 100.00 & 99.20 \\
& DistilBERT+LoRA & 92.31 & 99.49 & 99.73 & 96.15 & 100.00 & 97.86 \\
\hline
\multirow{10}{*}{\rotatebox{90}{Non-LLM Models}}
& CNN & 96.48 & 99.74 & 99.86 & 98.24 & 100.00 & 99.03 \\
& LSTM & 94.35 & 99.58 & 99.77 & 97.17 & 100.00 & 98.43 \\
& GRU & 96.98 & 99.78 & 99.88 & 98.49 & 100.00 & 99.17 \\
& Random Forest & 95.73 & 99.68 & 99.83 & 97.86 & 100.00 & 98.82 \\
& Decision Tree & 93.59 & 98.56 & 96.89 & 96.71 & 99.80 & 97.61 \\
& SVM & 92.34 & 99.43 & 99.69 & 96.17 & 100.00 & 97.85 \\
& XGBoost & 94.85 & 99.62 & 99.79 & 97.42 & 100.00 & 98.57 \\
& Logistic Regression & 68.22 & 97.64 & 98.75 & 84.11 & 100.00 & 89.92 \\
& KNN & 95.60 & 99.66 & 99.76 & 97.80 & 100.00 & 98.75 \\
& Naive Bayes & 69.85 & 85.05 & 63.14 & 78.06 & 78.06 & 66.22 \\
\Xhline{3\arrayrulewidth}
\end{tabular}
\end{table*}

\Ablack{In contrast with prior LLM applications, this work differs not only in its application but also in the nature of the input data and the deployment setting. The input to our model consists of the six  TCDR current waveforms captured before and after TCDR triggering. 
Each of the six dimensions represents a series of current measurements over 32 time steps. The data considered thus comprises a total of 192 features, which is far greater than the number of features used in 3 other works discussed.
More particularly, analyzing the approach of textualizing sequential data for LLM classification in cases where the amount of numerical information to be encoded in one input is very large, constitutes an angle that differs from the existing discourse. In this case, hundreds of features must be compacted into a natural language setting that fits within token limit specifications, presenting a unique constraint on prompt engineering. As a result, the formatting cannot make use of large linguistic sequences for contextualization, which further differs from the current literature. Especially since non-proprietary, locally deployable LLM models have greater input size restrictions, it is important to consider how effective contextualization of measurement data can be in such a context.
Furthermore, instead of relying on large, cloud-based models, we evaluate multiple lightweight, locally-deployable LLMs, DistilBERT, GPT-2, and DistilBERT+LoRA, capable of being hosted directly at the substation level. This setup improves real-time response feasibility and protects sensitive relay data from network-based exposure.}

\section{ Performance Evaluation}

 \subsubsection{LLM Evaluation Scenarios}

 \Ablack{The performance evaluation in this paper is conducted using the IEEE Power System Relaying Committee (PSRC) D6 test system \cite{kundur2007power}, modeled based on the IEC-61850 standard and simulated in the OPAL-RT HYPERSIM platform. 
\begin{figure*}[t!]
    \centering
    \includegraphics[width=1\linewidth]{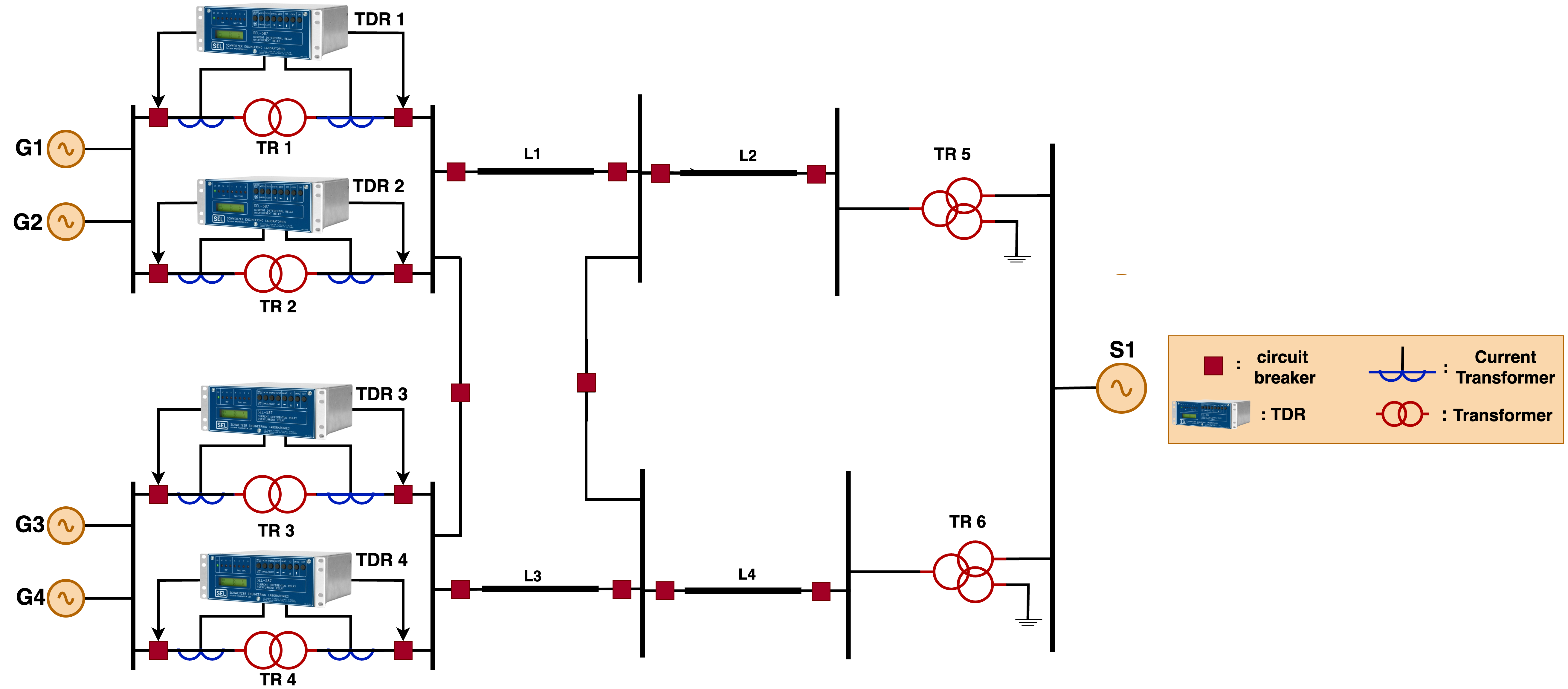}
    \caption{\Ablack{Test system.}}
    \label{fig:testtdr}
\end{figure*}
The system, illustrated in Fig, \ref{fig:testtdr}, 
includes four generators (G1–G4), and each generator feeds the system through a power transformer protected by a TCDR. Multiple current transformers (CT1–CT8) are leveraged, and merging units that transmit sampled current measurements as IEC-61850 Sampled Value (SV) packets to protective relays. The setup closely resembles real-world substations and allows for accurate modeling of protection logic and cyberattack scenarios.
Three fault types were simulated: single-phase-to-ground, two-phase-to-ground, and three-phase faults. For cyberattacks, three FDIA scenarios were considered: (i) direct injection of arbitrary current magnitudes, (ii) replay of captured fault data to mimic genuine events, and (iii) manipulation of transformer tap settings to create false differential conditions. Each simulation ran with varying generation levels (350–360 MW in 2 MW steps) and fault/attack events triggered between 1.00 and 1.02 seconds to introduce realistic temporal and operating variability.
The resulting dataset comprises $\approx$50,000 labeled samples. Each sample includes a 20 ms window of three-phase current data from both CT1 and CT2, sampled at 1600 Hz, forming an input matrix of shape (32, 6). Measurements are normalized to the range [0, 1] and converted into structured textual prompts. The dataset was split into 80\% for fine-tuning and 20\% for testing.}
\Ablack{We provide the full dataset and supplementary material used in this study at \url{https://github.com/jaafaris/LLMSmartGridTCDR}.}

\begin{figure*}[t!]
    \centering
    \includegraphics[width=1\linewidth]{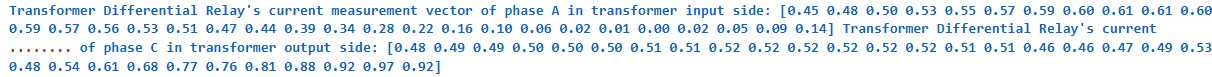}
    \caption{Example of a textualized FDIA sample generated using the structured prompt template.}
    \label{fig:textualized_example}
\end{figure*}

\begin{figure*}[t!]
    \centering
    \includegraphics[width=1\linewidth]{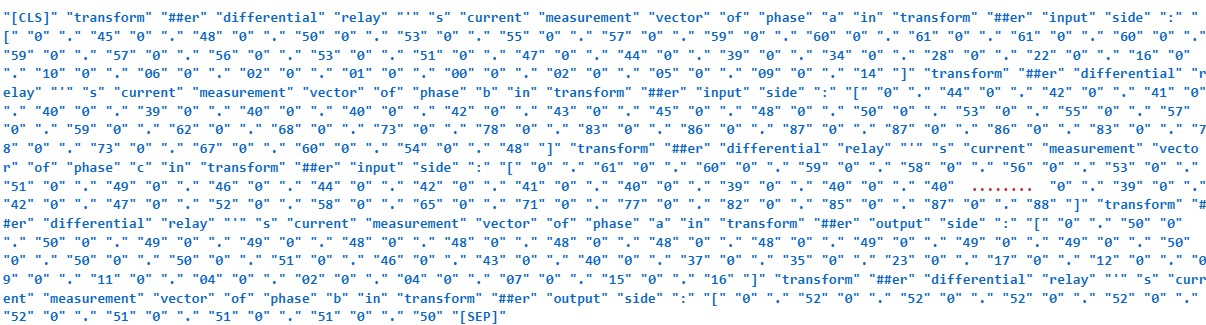}
    \caption{Tokenized representation of the textualized FDIA sample in Fig.~\ref{fig:textualized_example} using the HuggingFace \texttt{distilbert-base-uncased} tokenizer.}
    \label{fig:tokenized_example}
\end{figure*}

\Ablack{All three-phase current measurements from CT1 and CT2 are first normalized to the range [0,~1] and rounded to three decimal places to reduce token length while preserving sufficient numerical resolution. The normalized values are then formatted into fixed-length strings and embedded into a structured prompt template that preserves spatial (input-side vs.~output-side) and temporal ordering of the measurements. Tokenization is performed using the HuggingFace tokenizer associated with each model, producing an average sequence length of 512 tokens per sample, a number of tokens that can be supported by DistilBERT. The dataset is split into training and testing subsets using an 80/20 stratified split to maintain the original class distribution, and all experiments are performed with a fixed random seed of 42 to ensure reproducibility. Figures     \ref{fig:textualized_example} and  \ref{fig:tokenized_example} show an examples of an FDIA sample in the proposed prompt before and after tokenization.}


\subsubsection{Utilized LLMs' Settings}

We use Hugging Face's pre-trained DistilBERT model for sequence classification, \texttt{distilbert-base-uncased},
along with its associated DistilBERT tokenizer \cite{HuggingFace_DistilBERT}. DistilBERT is fine-tuned over 10 epochs using the Adam optimizer with a learning rate of \(2 \times 10^{-5}\), 10 logging steps, a batch size of 16, a weight decay of 0.01, and the best-performing model is selected for testing. 
Fine-tuning is conducted for approximately 1 hour in a Google Colab session using an A100 40GB GPU accelerator, twelve 6-core Intel Xeon CPUs (2.20GHz each), and 83.5GB of system RAM.
Additionally, two other LLMs are implemented: GPT-2 and a Low-Rank Adaptation (LoRA)-tuned variant of DistilBERT. 
The LoRA approach freezes most of the original model’s parameters and introduces a smaller number of trainable parameters, minimizing computational overhead. The LoRA+DistilBERT model is fine-tuned with a dropout rate of 0.05, \(\alpha = 32\), and \(r = 8\) \cite{HuggingFace_DistilBERT}.
 The GPT-2 model is fine-tuned on the same dataset using the GPT-2 tokenizer and the same training parameters \cite{HuggingFace_DistilBERT}.

\subsubsection{ Results Discussion}
Table~\ref{tab:weighted_metrics} depicts the testing results of DistilBERT and other implemented state-of-the-art models, including GPT-2, DistilBERT+LoRA, gated recurrent unit (GRU), convolutional neural network (CNN), long short-term memory (LSTM), support vector machine (SVM),  eXtreme Gradient Boosting (XGBoost), and  k-Nearest Neighbors (KNN).  
These models are evaluated using the metrics detected cyberattack (true positive) rate, accuracy, precision (macro-averaged),  recall (macro-averaged), specificity (true negative rate, and F1-Score (macro-averaged) \cite{scikit-learn-precision-recall}. 
Our results confirm that DistilBERT continues to outperform or match the performance of  several strong baselines across key metrics, while remaining competitive with the highest-performing models. These findings reinforce the conclusion that lightweight, locally deployable LLMs can achieve detection performance that is better or comparable to state-of-the-art learning-based approaches, while retaining operational feasibility in substation environments. In addition to accuracy, a key advantage of LLM-based classifiers such as DistilBERT lies in their interpretability. 
Attention mechanisms in LLMs provide valuable insight into decision-making, which blackbox models like GRU, CNN and LSTM
lack.
Using attention weights, we can visualize which segments of the input measurement sequence the model relied upon when making decisions, offering valuable insights for operators and enhancing model transparency.

In detail, DistilBERT is effective for detecting cyberattacks targeting TCDRs, detecting 97.62\% of cyberattacks that would have otherwise caused false tripping of the TCDR given its existing TCDR logic. In our results, DistilBERT fully preserves the TCDR’s fault detection accuracy.
Similarly, GPT-2 and DistilBERT+LoRA achieve 97.06\% and 92.31\%, respectively, in terms of detected cyberattacks.
While LoRA reduces training parameters, in this application, it limits the model's ability to fully adapt to the nuances and subtle patterns of the TCDR measurements necessary to detect cyberattacks, leading to  lower accuracy compared to full fine-tuning. This underscores the importance of full fine-tuning for tasks requiring high precision and domain-specific adaptation.
One can also observe that the performance of all three LLMs is either better or comparable to that of state-of-the-art DL and ML models, while Logistic Regression and Naive Bayes perform poorly, with a high percentage of undetected cyberattacks; 31.78\% and 30.15\%, respectively.
The strong performance of the three LLMs in this application can be attributed to their ability to capture complex contextual relationships across the textualized sequences of TCDR measurements through self-attention mechanisms. This underscores the potential of using LLMs for resilient cyberattack detection in TCDRs and similar problems.

\begin{figure}[t!]
    \centering
    \begin{minipage}{1\columnwidth} 
        \centering
        \includegraphics[width=1\linewidth]{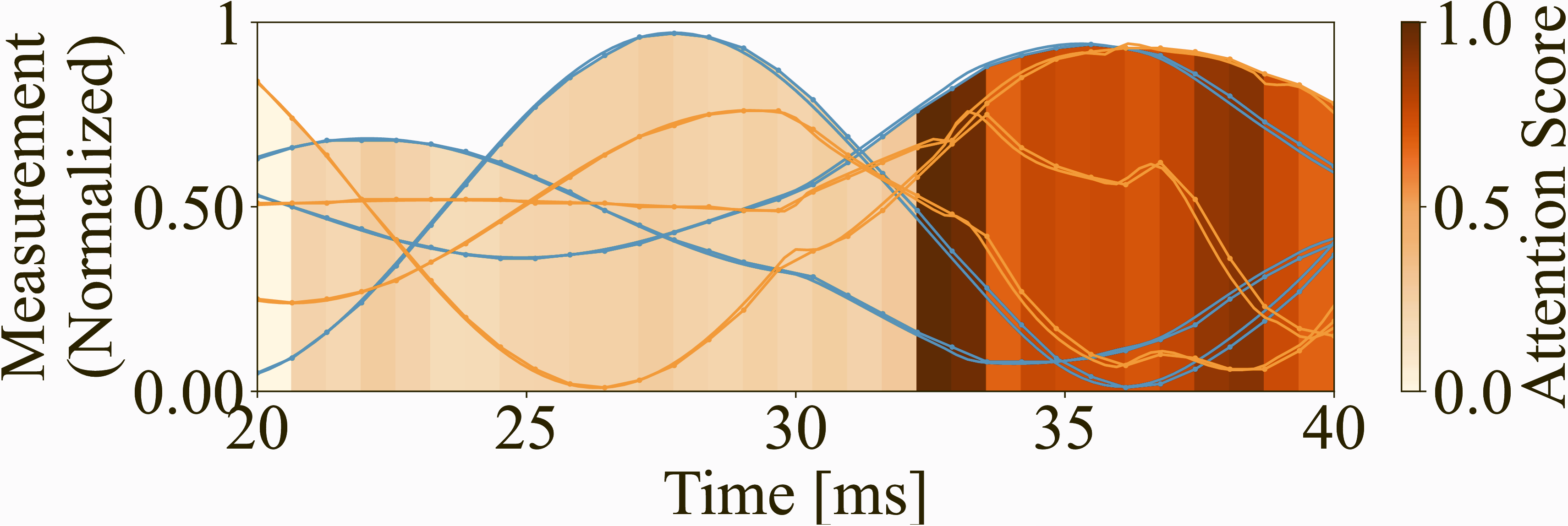}
    \end{minipage}\\[0.05cm]
    \caption{Attention weights assigned to TCDR measurements under a cyberattack.}
    \label{fig:measurements_with_attention}
\end{figure}

Figure \ref{fig:measurements_with_attention}, illustrates  attention scores assigned by our fine-tuned DistilBERT model for a detected cyberattack sample. The figure shows the TCDR normalized measurements for phase A to C on both the input (blue) and output (orange) sides.
Attention scores are overlaid on the measurements as a heatmap.  
It can be observed that regions with highest attention scores (darkest red regions) across all phases appear at $t\approx$33$-$38 ms, aligning with the time steps where the attack unfolds, providing insights into how the LLM detects cyberattacks.
\Ablack{Semantically, higher attention scores on a part of the sequence indicate greater relevance of that region to the model’s classification decision. In this context, the attention score at each time point reflects how strongly the LLM associates the measurement at that time with a cyberattack or a fault. Where the model has classified a cyberattack, these scores provide an operator with a numerical and visual heuristic indicating which time regions in the measurement data are most indicative of the potential cyberattack, and at what point the model begins to detect suspicious patterns. For example, a sudden rise in attention may correspond to a transient in the current measurements that is inconsistent with known fault characteristics, signalling to the practitioner when and why the model flags the scenario as an attack. Thus, the attention map not only aids in understanding the model’s decision but can also guide operators in investigating specific anomalies within the sequence. This interpretability feature allows protection engineers to validate the model’s decision.  This capability provides an additional advantage of the LLM-based approach compared to conventional deep learning models, which lack such inherent interpretability.}

\Ablack{All three investigated LLMs achieve consistently high detection performance across all evaluation scenarios, confirming the strong capability of LLMs to distinguish cyberattacks from genuine faults in TCDRs. While non-LLM models such as CNNs or GRUs can also perform well, they lack intrinsic interpretability. In contrast, the proposed LLM-based classifiers provide transparent decision reasoning through attention-based visualization of the most influential measurement segments, supporting operator trust and situational awareness. With the ongoing advances in lightweight LLM architectures, future versions are expected to offer even higher accuracy while retaining interpretability and suitability for locally deployable cyberattack detection in digital substations.}

\subsection{\Ablack{Performance Evaluation Under Complex Cyberattacks}}

\Ablack{To further assess the generalization capability of the proposed approach, we evaluate all considered models under a complex, combined attack scenario that combines a time-stamp attack (TSA) with an FDIA. In this scenario 
a TSA manipulates the time stamp of the TCDR's output-side measurements, introducing a 1 ms time delay between the input and output current waveforms of the TCDR,
immediately followed by an FDIA payload that attempts to mimic a genuine fault. Models are trained on the noise-free training set described in Section~III and tested on this perturbed holdout set without any retraining, thereby simulating an unforeseen adversarial condition at inference time.}

\Ablack{Table~\ref{tab:cmplx_attacks} reports the detected-cyberattack rates for LLM and non-LLM models under this worst-case scenario. The results show that LLMs achieve high detection performance (DistilBERT: 97.20\%, GPT-2: 96.64\%), comparable to the best non-LLM baselines: GRU: 97.24\%, CNN: 96.48\%. DistilBERT+LoRA shows slightly reduced detection in this setting, 92.02\%, reflecting the trade-offs introduced by parameter-efficient adaptation under severe perturbations. These findings indicate that LLMs effectively learn discriminative signatures that distinguish true physical faults, which follow power-system physical constraints, from malicious manipulations of relay measurements; in particular, the models leverage subtle cross-channel and temporal inconsistencies induced by the attack payload that are difficult for simple rule-based detectors to capture. While no single model is uniformly superior for every possible composite attack, the results confirm that locally deployable LLMs are capable of accurately detecting complex cyberattacks on protective relays, supporting their practical viability as an additional cyber-defense layer.}

\subsection{\Ablack{Impact of Measurement Noise}}

\Ablack{In this subsection, we further evaluate model robustness to unforeseen measurement noise by testing all trained models on holdout sets corrupted with additive white Gaussian noise at several signal-to-noise ratio (SNR) levels, including 45, 40, 35, and 30 dB. This experiment models non-ideal measurement conditions that may arise from lower-grade current transformers or adverse measurement chain conditions. Recall that all models were trained earlier on noise-free FDIAs and faults. These models are now evaluated without retraining.}

\Ablack{Table~\ref{tab:SNR} summarizes the models' accuracy under different SNRs. At high SNRs (45–40 dB) nearly all models maintain accuracy above 99\%. At moderate noise, 35 dB, modest differences begin to appear: DistilBERT retains 96.19\% accuracy, while GPT-2 and DistilBERT+LoRA remain above 99\% in our tests. At the lowest tested SNR, 30 dB, accuracy divergence becomes more pronounced; DistilBERT drops to 89.61\%, whereas some architectures, such as GRU/CNN/XGBoost variants, remain close to 99\% in this dataset and configuration. These observations indicate two conclusions: (i) when measurement noise is small (realistically expected in modern substations using high-grade measurement devices), LLMs and conventional models perform acceptably and comparably; (ii) when noise becomes substantial, detection performance may degrade for some LLM configurations.  This slight degradation in performance can be mitigated in practice by ensuring the use of only  high-precision measurement devices in modern substations.}

\begin{table}[t!]
\centering
\caption{Performance Under Complex Cyberattacks}
\label{tab:cmplx_attacks}
\renewcommand{\arraystretch}{1.25}
\begin{tabular}{l  l c c}
\toprule
 & \textbf{Model} & \textbf{Detected Complex Cyberattacks (\%)} \\
\midrule
\multirow{3}{*}{\rotatebox{90}{\textbf{LLMs}}} 
 & DistilBERT   & 97.20 \\
 & DistilBERT+LORA   & 92.02 \\
 & GPT-2        & 96.64  \\
\midrule
\multirow{10}{*}{\rotatebox{90}{\textbf{Non-LLM Models}}} 
 & CNN                  & 96.48  \\
& LSTM                 & 94.10 \\
 & GRU                  & 97.24 \\
 & Decision Tree        & 88.82  \\
 
 & KNN                  & 95.60  \\
 & Logistic Regression  & 67.34  \\
 
 & Random Forest        & 96.23  \\
 & SVM                  & 92.71  \\
 & XGBoost              & 91.33  \\
 & Naive Bayes          & 69.72  \\
\bottomrule
\end{tabular}
\end{table}

\begin{table}[t!]
\centering
\caption{Models' Accuracy (\%) Under Measurement Noise with Different SNRs}
\label{tab:SNR}
\renewcommand{\arraystretch}{1.25}
\begin{tabular}{c l c c c c}
\Xhline{3\arrayrulewidth}
 & \textbf{Model} & \textbf{45 dB} & \textbf{40 dB} & \textbf{35 dB} & \textbf{30 dB} \\
\Xhline{3\arrayrulewidth}
\multirow{3}{*}{\rotatebox{90}{\textbf{LLMs}}}
& DistilBERT & 99.52 & 98.79 & 96.19 & 89.61 \\
 & DistilBERT + LoRA & 99.47 & 99.43 & 99.42 & 99.23 \\
 & GPT-2 & 99.74 & 99.68 & 99.33 & 97.82 \\
 [3pt]
\Xhline{2\arrayrulewidth}
\multirow{10}{*}{\rotatebox{90}{\textbf{Non-LLM Models}}}
 & CNN & 99.74 & 99.73 & 99.73 & 99.73\\
 & LSTM & 99.58 & 99.57 & 99.57 & 99.57\\
 & GRU & 99.77 & 99.77 & 99.77 & 99.77\\
 & Random Forest & 99.70 & 99.70 & 99.68 & 99.68 \\
 & Decision Tree & 97.92 & 95.93 & 91.62 & 83.85\\
 & SVM & 99.43 & 99.43 & 99.43 & 99.42\\
 & XGBoost & 99.59 & 99.56 & 99.53 & 99.39\\
 & Logistic Regression & 97.63 & 97.63 & 97.63 & 97.63\\
 & KNN & 99.67 & 99.67 & 99.66 & 99.66\\
 & Naive Bayes & 85.06 & 85.05 & 85.00 & 85.00\\
\Xhline{3\arrayrulewidth}
\end{tabular}
\end{table}

\subsection{\Ablack{Sensitivity to Prompt Variations}}
\Ablack{To analyze the sensitivity of the proposed LLM-based framework to prompt formulation, an ablation study was conducted using three controlled prompt variants alongside the baseline configuration, as summarized in Table~\ref{tab:prompt_variants_summary}. Each variant modifies one linguistic or structural aspect of the prompt while keeping all numerical content identical. Variant~1 rephrases the English descriptions into shorter and more direct sentences to test the impact of linguistic brevity. Variant~2 alters the structural order of measurements from the baseline sequence
($\mathrm{A}_{\mathrm{in}}, \mathrm{B}_{\mathrm{in}}, \mathrm{C}_{\mathrm{in}}, 
\mathrm{A}_{\mathrm{out}}, \mathrm{B}_{\mathrm{out}}, \mathrm{C}_{\mathrm{out}}$)
to an alternating layout 
($\mathrm{A}_{\mathrm{in}}, \mathrm{A}_{\mathrm{out}}, \mathrm{B}_{\mathrm{in}}, 
\mathrm{B}_{\mathrm{out}}, \mathrm{C}_{\mathrm{in}}, \mathrm{C}_{\mathrm{out}}$),
examining whether token positioning influences inference. Variant~3 removes explicit identifiers such as phase labels and input/output side, thereby testing the model’s reliance on explicit domain semantics for context. 
Our results presented in Tables~\ref{tab:distilbert_ablation_percent}, \ref{tab:distilbert_LoRA_ablation_percent} and \ref{tab:GPT2_ablation_percent} indicate that both DistilBERT and DistilBERT+LoRA exhibit strong robustness to these variations. Across all three prompt types, model accuracy remains above 99\%, and the cyberattack detection rate fluctuates within a narrow 1–2\% range relative to the baseline. This stability demonstrates that the models primarily learn underlying numerical relationships between current waveforms rather than depending on exact wording or prompt order.
However, for two of the three models (DistilBERT and DistilBERT+LoRA), omitting contextual identifiers in Variant~3 led to a modest decrease in recall,
confirming that explicit semantic cues, such as phase and side information, help the LLM better capture spatial–temporal measurement dependencies.}

\Ablack{These findings confirm that the proposed textualization strategy used in this work is effective for representing transformer relay measurements. 
While the current prompt template is not claimed to be globally optimal, it constitutes a robust and effective formulation. Other semantically equivalent prompts could be adopted in future studies without significantly affecting detection accuracy, offering flexibility for adapting the framework to other protective relays and data modalities.}

\begin{table*}[!t]
\centering
\caption{Summary of Investigated Prompt Template Variants}
\label{tab:prompt_variants_summary}
\renewcommand{\arraystretch}{1.25}
\begin{tabular}{p{1.6cm} p{8cm} p{4.5cm}}
\Xhline{3\arrayrulewidth}
\textbf{Prompt} & \textbf{Prompt Template Description} & \textbf{Notes} \\
\Xhline{3\arrayrulewidth}
\textcolor{white}{.} \newline 
\textcolor{white}{.} \newline 
\textbf{Baseline} &
Transformer Differential Relay's current measurement vector of phase A in transformer input side: [values] \newline
Transformer Differential Relay's current measurement vector of phase B in transformer input side: [values] ...  \newline
Transformer Differential Relay's current measurement vector of phase C in transformer output side: [values] &
 Baseline prompt; introduces the main information about the relay measurements in a simple structured manner
\\
\midrule
\textcolor{white}{.} \newline 
\textcolor{white}{.} \newline 
\textbf{Variant 1 \newline (Phrasing)} &
Differential relay phase A current measurements at transformer input side: [values] \newline
Differential relay phase B current measurements at transformer input side: [values] ... \newline
Differential relay phase C current measurements at transformer output side: [values] &
Tests sensitivity to prompt phrasing and 
by modifying the phrasing while keeping all domain semantics unchanged. 
\\
\midrule
\textcolor{white}{.} \newline 
\textcolor{white}{.} \newline 
\textbf{Variant 2 \newline (Structure)} &
Transformer Differential Relay's current measurement vector of phase A in transformer input side: [values] \newline
Transformer Differential Relay's current measurement vector of phase A in transformer output side: [values] \newline
Transformer Differential Relay's current measurement vector of phase B in transformer input side: [values] ... &
Changes prompt structure by rearranging measurements order to asses whether token ordering affects inference.
\\
\midrule
\textcolor{white}{.} \newline 
\textbf{Variant 3 \newline (Key Info \newline Removed)} &
Transformer Differential Relay's current measurement vectors are: 
[numerical values of phase A at transformer input side];
[numerical values of phase B at transformer input side]; ... \newline
[numerical values of phase C at transformer output side]; &
Removes key information phase and side identifiers to assess reliance on explicit contextual cues. 
\\
\Xhline{3\arrayrulewidth}
\end{tabular}
\end{table*}

\begin{table}[!t]
\centering
\caption{
Performance of DistilBERT Under Prompt Variants
}
\label{tab:distilbert_ablation_percent}
\renewcommand{\arraystretch}{1.25}
\begin{tabular}{lcccc}
\toprule
\textbf{Metric}
& \textbf{Baseline}
& \textbf{Variant 1}
& \textbf{Variant 2}
& \textbf{Variant 3} \\
\midrule
DetectedAttacks (\%)  & 97.62 & 95.51 & 94.56 & 94.56 \\
Accuracy (\%)                    & 99.84 & 99.71 & 99.63 & 99.63 \\
Precision (\%)                   & 100.00 & 100.00 & 100.00 & 100.00 \\
Recall (\%)                      & 98.81 & 98.00 & 97.00 & 97.00 \\
Specificity (\%)                 & 100.00 & 100.00 & 100.00 & 100.00 \\
F1-Score (\%)                    & 99.36 & 99.00 & 99.00 & 99.00 \\
\bottomrule
\end{tabular}
\end{table}

\begin{table}[!t]
\centering
\caption{
Performance of DistilBERT+LORA Under Prompt Variants
}
\label{tab:distilbert_LoRA_ablation_percent}
\renewcommand{\arraystretch}{1.25}
\begin{tabular}{lcccc}
\toprule
\textbf{Metric}
& \textbf{Baseline}
& \textbf{Variant 1}
& \textbf{Variant 2}
& \textbf{Variant 3} \\
\midrule
DetectedAttacks (\%) & 92.31 & 89.88 & 92.56 & 87.99\\
Accuracy (\%) & 99.49 & 99.28 & 99.49 & 99.20  \\
Precision  (\%) & 99.73 & 99.62 & 99.73 & 99.57  \\
Recall (\%) & 96.15 & 94.94 & 96.28 & 93.99 \\
Specificity (\%) & 100.00 & 100.00 & 100.00 & 100.00 \\
F1-Score (\%) & 97.86 & 97.14 & 97.93 & 96.59\\
\bottomrule
\end{tabular}
\end{table}

\begin{table}[!t]
\centering
\caption{
Performance of GPT2 Under Prompt Variants
}
\label{tab:GPT2_ablation_percent}
\renewcommand{\arraystretch}{1.25}
\begin{tabular}{lcccc}
\toprule
\textbf{Metric}
& \textbf{Baseline}
& \textbf{Variant 1}
& \textbf{Variant 2}
& \textbf{Variant 3} \\
\midrule
DetectedAttacks  (\%) & 97.06 & 95.95 & 95.81 & 95.81 \\
Accuracy (\%)  & 99.80 & 99.73 & 99.72 & 99.72 \\
Precision  (\%) & 99.90 & 99.86 & 99.85 & 99.85 \\
Recall (\%) & 98.53 & 97.97 & 97.91 & 97.91 \\
Specificity (\%) &  100.00 & 100.00 & 100.00 & 100.00 \\
F1-Score (\%) & 99.20 & 98.89 & 98.86 & 98.86\\
\bottomrule
\end{tabular}
\end{table}

\subsection{Time Complexity of Proposed Solution} 

To ensure that the proposed LLM-based framework operates within the stringent timing requirements of TCDRs, we evaluate the inference time of the trained DistilBERT model. Using 10{,}000 test cases of both faults and cyberattacks, executed on a workstation equipped with an Intel Core i9 processor and an NVIDIA GeForce RTX~4060 GPU, we measure an average inference time of 5.39~ms per sample. This corresponds to approximately 0.32 cycles at a 60~Hz system frequency. Given that TCDRs are expected to detect and trip within 2–3 cycles \cite{SEL_inc}, the proposed approach satisfies real-time protection requirements while retaining its detection performance advantages. \Ablack{This real-time performance result reinforces the technological contribution of the framework, demonstrating that lightweight large language models can meet  protective-relay timing constraints and operate entirely within substation environments without reliance on external compute resources.}

\section{Discussion}

This is while the proposed LLM-based detection framework substantially enhances the ability of TCDRs to identify cyberattacks, its deployment in operational substations must be approached with careful consideration of potential new cybersecurity risks. In particular, network-connected models can introduce additional attack surfaces, including adversarial machine learning attacks, model poisoning, and Denial-of-Service (DoS) attempts. In adversarial machine learning scenarios, carefully crafted inputs could be designed to exploit the LLM’s learned representations, forcing it to misclassify malicious activity as benign. Model poisoning could occur if malicious data is introduced into the model’s training or update processes, thereby degrading its detection capabilities over time. DoS attacks, on the other hand, may aim to overload the inference pipeline, delaying or preventing timely protective action.
To mitigate these risks, the LLM should be deployed entirely on-device within an isolated, air-gapped substation network, ensuring that sensitive measurement data never leaves the local environment. All incoming measurement streams can be cryptographically authenticated to guarantee data integrity and origin, while rate-limiting and access-control mechanisms can restrict resource usage to authorized processes only. Furthermore, adversarial robustness measures, such as adversarial training, input sanitization, and anomaly detection, can provide additional resilience against manipulation attempts. These mitigation strategies are particularly important to preserve the high-speed operational requirements of TCDRs while extending their protection capabilities into the cybersecurity domain.

Moreover, the proposed LLM-based detection framework may introduce higher computational demands than traditional protection schemes, particularly during the training and fine-tuning phases, which benefit from GPU acceleration and adequate memory capacity. However, unlike conventional protective relays, this framework is not designed to replace primary protection functions; instead, it complements them by detecting cyberattacks that could otherwise be misinterpreted as legitimate faults. Our experiments show that fine-tuning can be performed on commercial computing systems, and the lightweight LLM architectures employed, such as DistilBERT, require significantly less computational power during inference, enabling sub-cycle decision times. Additionally, the ongoing digitalization of substations has led to the adoption of edge-computing nodes for monitoring and automation. These existing resources could be leveraged to deploy the LLM without substantial additional hardware investment. The incremental cost of deployment must be weighed against the improved cybersecurity resilience, and future work will focus on optimizing computational efficiency while quantifying this cost–benefit trade-off.

\Ablack{While the present framework focuses on detecting FDIAs targeting sampled current measurements of TCDRs, power system operators should be aware of other emerging cyber threats that LLMs themselves can be prone to. These include vulnerabilities such as backdoor attacks, adversarial machine learning attacks, model poisoning, and supply-chain compromises, which may affect pretrained models provided by third-party vendors. These threats introduce fundamentally different attack surfaces at the model and system level and require complementary defenses such as adversarial training, secure update pipelines, and access-control mechanisms. As such, addressing these attacks remains an important direction for future work, and it is important for power system operators to remain aware of these risks as LLM-based tools become more common in substations.}

\Ablack{Overall, the results of this paper confirm that lightweight, locally deployable LLMs achieve performance that is better or comparable to existing deep learning models, while also providing built-in interpretability through attention visualization. Importantly, these models are fine-tuned on commercial hardware without the need for large-scale training infrastructure, making them a practical and secure choice for substation deployment. Looking ahead, with the rapid evolution of LLMs, future lightweight variants are expected to achieve even higher detection accuracy while preserving their interpretability. Building on the framework presented in this paper, such models can further strengthen cyberattack detection capabilities in digital substations in a practical and operator-trusted manner.}

\section{Conclusion and Future Work}

This paper demonstrated the potential of LLMs for detecting cyberattacks on TCDRs in modern power systems. Our results showed that properly adapted LLMs, namely DistilBERT, GPT-2, and DistilBERT+LoRA, can capture complex relationships in textualized TCDR measurements, effectively detecting cyberattacks while preserving relay dependability. 
Overall, the results of this paper confirm that locally deployable LLMs such as DistilBERT can achieve performance comparable to or exceeding existing learning-based detection methods for TCDRs, with LLMs having the benefit of better interpretability.
\Ablack{Furthermore, the proposed LLM-based approach demonstrates robustness under complex combined attack scenarios and measurement noise, exhibits stable performance across different prompt variants, and has an inference latency below 5.5 ms on standard commercial hardware.}
The success of multiple LLMs in securing TCDRs highlights the promise of LLMs for enhancing smart grid cybersecurity. With the rapid pace of advancements in large language models, future generations of LLMs, offering improved architectures, training methods, and efficiency, are expected to deliver even higher detection accuracy. Exploring these next-generation models for substation protection is a promising direction for future work. Future work will also investigate the design of adversarially robust LLM architectures and designing tailored cybersecurity hardening measures to ensure safe deployment in substation environments. \Ablack{Future researchers can also investigate the vulnerability of the proposed LLM-based approach to emerging attacks directly targeting LLMs, developing proper solutions,} and \Ablack{investigating the impact of prompt engineering/optimization on the performance of the proposed framework}.

\section*{Acknowledgment}

This work was supported by the Natural Sciences and Engineering Research Council of Canada (NSERC) through the Discovery Grants Program

\bibliographystyle{IEEEtran} 

\bibliography{TII-Articles-LaTeX-template/Manuscript}

\begin{thebibliography}{10}
\providecommand{\url}[1]{#1}
\csname url@samestyle\endcsname
\providecommand{\newblock}{\relax}
\providecommand{\bibinfo}[2]{#2}
\providecommand{\BIBentrySTDinterwordspacing}{\spaceskip=0pt\relax}
\providecommand{\BIBentryALTinterwordstretchfactor}{4}
\providecommand{\BIBentryALTinterwordspacing}{\spaceskip=\fontdimen2\font plus
\BIBentryALTinterwordstretchfactor\fontdimen3\font minus \fontdimen4\font\relax}
\providecommand{\BIBforeignlanguage}[2]{{%
\expandafter\ifx\csname l@#1\endcsname\relax
\typeout{** WARNING: IEEEtran.bst: No hyphenation pattern has been}%
\typeout{** loaded for the language `#1'. Using the pattern for}%
\typeout{** the default language instead.}%
\else
\language=\csname l@#1\endcsname
\fi
#2}}
\providecommand{\BIBdecl}{\relax}
\BIBdecl

\bibitem{10065529}
S.~M.~S. Hussain, M.~A. Aftab, S.~M. Farooq, I.~Ali, T.~S. Ustun, and C.~Konstantinou, ``An effective security scheme for attacks on sample value messages in iec 61850 automated substations,'' \emph{IEEE Open Access Journal of Power and Energy}, vol.~10, pp. 304--315, 2023.

\bibitem{sanh2019distilbert}
\BIBentryALTinterwordspacing
V.~Sanh, L.~Debut, J.~Chaumond, and T.~Wolf, ``Distilbert, a distilled version of {BERT:} smaller, faster, cheaper and lighter,'' \emph{CoRR}, vol. abs/1910.01108, 2019. [Online]. Available: \url{http://arxiv.org/abs/1910.01108}
\BIBentrySTDinterwordspacing

\bibitem{zhao2024large}
T.~Zhao, A.~Yogarathnam, and M.~Yue, ``A large language model for determining partial tripping of distributed energy resources,'' \emph{IEEE Transactions on Smart Grid}, vol.~16, no.~1, pp. 437--440, 2025.

\bibitem{zaboli2024chatgpt}
A.~Zaboli, S.~L. Choi, T.-J. Song, and J.~Hong, ``Chatgpt and other large language models for cybersecurity of smart grid applications,'' in \emph{2024 IEEE Power \& Energy Society General Meeting (PESGM)}, 2024, pp. 1--5.

\bibitem{11245588}
H.~Chen, J.~Chen, Y.~Chai, W.~Guo, C.~Jia, B.~Yang, and Z.~Xin, ``Scene-aware non-intrusive load monitoring using large language models,'' \emph{IEEE Transactions on Smart Grid}, pp. 1--1, 2025, doi: 10.1109/TSG.2025.3632609.

\bibitem{SEL_inc}
\BIBentryALTinterwordspacing
{Schweitzer Engineering Laboratories, Inc.}, ``{SEL}-{T400L} time-domain line protection,'' WA, USA, 2021, accessed: 2025-06-13. [Online]. Available: \url{https://selinc.com/api/download/116461/}
\BIBentrySTDinterwordspacing

\bibitem{khaw2020deep}
Y.~M. Khaw, A.~Abiri~Jahromi, M.~F.~M. Arani, S.~Sanner, D.~Kundur, and M.~Kassouf, ``A deep learning-based cyberattack detection system for transmission protective relays,'' \emph{IEEE Transactions on Smart Grid}, vol.~12, no.~3, pp. 2554--2565, 2021.

\bibitem{kundur2007power}
P.~Kundur, ``Power system stability,'' \emph{Power system stability and control}, vol.~10, no.~1, pp. 7--1, 2007.

\bibitem{HuggingFace_DistilBERT}
\BIBentryALTinterwordspacing
{Hugging Face}, ``Distil{BERT}: Smaller, faster, cheaper, lighter,'' accessed: 2025-08-09. [Online]. Available: \url{https://huggingface.co/docs/transformers/en/model_doc/distilbert}
\BIBentrySTDinterwordspacing

\bibitem{scikit-learn-precision-recall}
\BIBentryALTinterwordspacing
{Scikit-learn Developers}. sklearn.metrics.precision\_recall\_fscore\_support -- scikit-learn documentation. Accessed: 2025-08-09. [Online]. Available: \url{https://scikit-learn.org/stable/modules/generated/sklearn.metrics.precision_recall_fscore_support.html}
\BIBentrySTDinterwordspacing

\end{thebibliography}

\end{document}